\RequirePackage[running]{lineno}
\documentclass[aps,prl,nofootinbib,twocolumn,showpacs,superscriptaddress,letterpaper,amsmath,amssymb]{revtex4-1}

\pdfoutput=1

\usepackage{graphicx}
\DeclareGraphicsExtensions{.pdf}

\usepackage{dcolumn}
\usepackage{bm}
\usepackage{feynmf}
\usepackage{slashed}
\usepackage{multirow}
\usepackage{placeins}

\usepackage{atlasphysics}

\usepackage{subfigure}
\usepackage{units}

\def\gsim{\lower0.5ex\hbox{$\:\buildrel >\over\sim\:$}}
\def\lsim{\lower0.5ex\hbox{$\:\buildrel <\over\sim\:$}}

\def\metv{$\textbf{\textit{E}}_{\rm T}^{\rm miss}$}
\def\missET {E_{\textrm{T}}^{\textrm{miss}}}
\def \Mstar{M_{*}}

\def\MET{E_{\rm T}^{\rm miss}}

\DeclareGraphicsExtensions{.pdf,.png,.jpg,.jpeg,.eps}

\newcommand\wgamma{\ensuremath{W(\rightarrow \ell\nu) + \gamma}}
\newcommand\zgamma{\ensuremath{Z(\rightarrow \ell\ell) + \gamma}}
\newcommand\gammaznunu{\ensuremath{Z(\rightarrow \nu\nu) + \gamma}}
\newcommand\wzjet{\ensuremath{W/Z + {\rm jet}}}
\newcommand\gammajets{\ensuremath{\gamma + {\rm jet}}}

\newcommand{\rom}[1]{\uppercase\expandafter{\romannumeral #1\relax}}

\newcommand\jet{\mathrm{jet}}

\newcommand{\pythia}{\textsc{Pythia}}
\newcommand{\pythiaeight}{\textsc{Pythia 8}}
\newcommand{\powheg}{\textsc{Powheg}}
\newcommand{\sherpa}{\textsc{Sherpa}}

\newcommand{\madgraph}{\textsc{Madgraph}}
\newcommand{\herwig}{\textsc{Herwig}}
\newcommand{\acermc}{\textsc{AcerMC}}
\newcommand{\mcatnlo}{\textsc{MC@NLO}}
\newcommand{\jimmy}{\textsc{Jimmy}}

\usepackage{preprintcover} 

\usepackage[hyperindex,breaklinks]{hyperref}

\PreprintCoverPaperTitle{Search for new phenomena in events with a photon and missing transverse momentum in $pp$ collisions at $\sqrt{s}=8$~TeV with the ATLAS detector}  

\PreprintIdNumber{CERN-PH-EP-2014-245}  

\PreprintCoverAbstract{Results of a search for new phenomena in events with an energetic photon and large missing transverse momentum with  the ATLAS experiment at the LHC are reported.  Data were collected in proton--proton collisions at a center-of-mass energy of 8 TeV and correspond to an integrated  luminosity of 20.3 fb$^{-1}$.  The observed data are well described by the expected Standard Model backgrounds.   The expected (observed) upper limit on the fiducial cross section for the production of events with a photon and large missing transverse momentum is 6.1 (5.3) fb at 95\% confidence level. Exclusion limits are presented on models of new phenomena with large extra spatial dimensions, supersymmetric quarks, and direct pair production of dark--matter candidates.}  

\PreprintJournalName{Phys.~Rev.~D}

\begin{document}

\thispagestyle{empty}
\vspace*{-3.5cm}

\vspace{0.5in}

\title{Search for new phenomena in events with a photon and missing transverse momentum in $pp$ collisions at $\sqrt{s}=8$~TeV with the ATLAS detector}

\begin{center}
\begin{abstract}
Results of a search for new phenomena in events with an energetic photon and large missing transverse momentum with  the ATLAS experiment at the LHC are reported.  Data were collected in proton--proton collisions at a center-of-mass energy of 8 TeV and correspond to an integrated  luminosity of 20.3 fb$^{-1}$.  The observed data are well described by the expected Standard Model backgrounds.   The expected (observed) upper limit on the fiducial cross section for the production of events with a photon and large missing transverse momentum is 6.1 (5.3) fb at 95\% confidence level. Exclusion limits are presented on models of new phenomena with large extra spatial dimensions, supersymmetric quarks, and direct pair production of dark--matter candidates.
\end{abstract}
\end{center}

\author{ATLAS Collaboration}


\date{ \today}
\maketitle

\section{I. Introduction}

Events that contain a high-momentum photon and large missing transverse momentum (referred to as $\gamma+\missET$) constitute a low-background sample that provides powerful sensitivity to some models of new phenomena~\cite{Aaltonen:2008hh,Abazov:2008kp,Abbiendi:2000hh,Achard:2003tx,Abdallah:2003np,
Aad:2012fw, Chatrchyan:2012tea}.  Theories with large extra spatial dimensions (LED), presence of dark--matter (DM), or supersymmetric partners of the quarks (squarks) in a compressed mass spectrum scenario predict the production of $\gamma+\missET$ events in $pp$ collisions beyond Standard Model (SM) expectations.

 The model of LED proposed by Arkani-Hamed, Dimopoulos, and Dvali~\cite{ArkaniHamed:1998rs} (ADD) aims to solve the hierarchy problem by hypothesizing the existence of $n$ additional spatial dimensions of size $R$, leading to a new fundamental scale $M_{\rm D}$ related to the Planck mass,  $M_{\rm Planck}$, through $M_{\rm Planck}^2 \approx M_{\rm D}^{2+{\rm n}}R^{\rm n}$.  If these dimensions are compactified, a series of massive graviton modes results.  These gravitons may be invisible to the ATLAS detector, but if the graviton is produced in association with a photon, the detector signature is a $\gamma+\missET$ event, as illustrated in  Fig.~\ref{fig:FeynmanADD}.

Although the presence of  DM is well established~\cite{Bertone:2004pz}, its possible particle nature remains a mystery.  A popular candidate is a weakly interacting massive particle (WIMP) denoted $\chi$, which has an interaction strength with SM particles at the level of the weak interaction.    If the WIMPs interact with quarks via a heavy mediator, they could be pair produced in collider events.  The $\chi\bar{\chi}$ pair would be invisible, but $\gamma+\missET$ events can be produced  via radiation of an initial-state photon in $q\bar{q}\chi\bar{\chi}$ interactions~\cite{Goodman:2010ku}.

As observations so far do not provide strong constraints on the nature of the WIMPs and the theoretical framework to which they belong, it is particularly interesting to study model-independent effective field theories (EFT) with various forms of interaction between the WIMPs and the Standard Model particles \cite{Goodman:2010ku}. In this framework, the mediator is effectively integrated out from the propagator and the production mechanism at the LHC energy scale is considered as a contact interaction, as illustrated in  Fig.~\ref{fig:FeynmanEFT}.  Several EFT operators for which the WIMP is a Dirac fermion are used as a representative set following the nomenclature of Ref.~\cite{Goodman:2010ku}: D5 (vector), D8 (axial vector), and D9 (tensor).  The interactions of SM and DM particles are described by two parameters: the DM particle mass $m_\chi$ and the suppression scale ($\Mstar$) of the heavy mediator. In an ultraviolet complete theory, the contact interaction would be replaced by an interaction via an explicit mediator $V$; the suppression scale is linked to the mediator mass $m_V$ by the relation $\Mstar=m_V/\sqrt{g_fg_\chi}$, where $g_f$ and $g_\chi$ represent the coupling factors of the mediator to SM particles and WIMPs, respectively.
However, as the typical momentum transfer in LHC collisions can reach the scale of the microscopic interaction, it is also crucial to probe specific models that involve the explicit production of the intermediate state, as shown in Fig.~\ref{fig:FeynmanLM}. In this case, the interaction is effectively described by four parameters:
$m_\chi$, $m_V$, the width of the mediator $\Gamma$, and the overall coupling $\sqrt{g_fg_\chi}$. In this paper, both the EFT approach presented in Ref.~\cite{Goodman:2010ku} and a specific model with a $Z'$-like mediator~\cite{LMWIMP} are considered.

An alternative DM model hypothesizes interactions between the WIMPs and SM gauge bosons~\cite{schannelWIMP}. The effective coupling to different bosons is parametrized by the coupling strengths $k_1$ and $k_2$, which control the strength of the coupling to the U(1) and SU(2) gauge sectors of the SM, respectively.   In this model, dark--matter production proceeds via $pp\rightarrow\gamma + X \rightarrow\gamma\chi\bar{\chi} + X'$, requiring no initial-state radiation, as shown in Fig.~\ref{fig:FeynmanFermi}. This model can also be used to describe the peak observed in the Fermi-LAT data~\cite{Weniger:2012tx}, allowing a direct comparison of Fermi and ATLAS data in the same parameter space.

Supersymmetry~\cite{Miyazawa:1966,Ramond:1971gb,Golfand:1971iw,Neveu:1971rx,Neveu:1971iv,Gervais:1971ji,Volkov:1973ix,Wess:1973kz,Wess:1974tw} postulates the existence of a new supersymmetric partner for each SM particle,  differing by half a unit of spin from, but with gauge coupling identical to, those of their SM counterparts.  Collisions of protons could result in pair production of squarks, $\tilde{q}$, which could decay to a SM quark and a neutralino $\tilde{\chi}_1^0$; the neutralino is assumed to be stable in  $R$-parity-conserving models~\cite{rparity}.   If the mass difference $m_{\tilde{q}}-m_{\tilde{\chi}_1^0}$ is small, the SM quarks would have very low momentum and would therefore not be reconstructed as jets.   Again, the radiation of a photon either from an initial-state quark or an intermediate squark would result in $\gamma+\missET$ events, as shown in Fig.~\ref{fig:FeynmanSUSY}.

\begin{figure}[htbp]\centering
  \includegraphics[width=0.5\linewidth]{{{fig_01}}}
  \caption{Graviton (G) production in models of large extra dimensions.}
  \label{fig:FeynmanADD}
\end{figure}

\begin{figure}[htbp]\centering
  \includegraphics[width=0.5\linewidth]{{{fig_02}}}
  \caption{Production of pairs of dark--matter particles ($\chi\bar{\chi}$) via an effective four-fermion $q\bar{q}\chi\bar{\chi}$ vertex.}
  \label{fig:FeynmanEFT}
\end{figure}

\begin{figure}[htbp]\centering
  \includegraphics[width=0.5\linewidth]{{{fig_03}}}
  \caption{Production of pairs of dark--matter particles ($\chi\bar{\chi}$) via an explicit $s$-channel mediator, V.}
  \label{fig:FeynmanLM}
\end{figure}

\begin{figure}[htbp]\centering
  \includegraphics[width=0.5\linewidth]{{{fig_04}}}
  \caption{Production of pairs of dark--matter particles ($\chi\bar{\chi}$) via an effective  $\gamma\gamma\chi\bar{\chi}$ vertex.}
  \label{fig:FeynmanFermi}
\end{figure}

\begin{figure}[htbp]\centering
  \includegraphics[width=0.5\linewidth]{{{fig_05}}}
  \caption{Pair production of squarks ($\tilde{q}$), followed by decay into quarks and neutralinos ($\tilde{\chi}_1^0$). The photon may also be radiated from the squarks or final-state quarks.}
  \label{fig:FeynmanSUSY}
\end{figure}

 The ATLAS~\cite{Aad:2012fw} and CMS~\cite{Chatrchyan:2012tea} collaborations have reported limits on various models of new physics based on searches for an excess in $\gamma+\missET$ events using $pp$ collisions at a center-of-mass energy $\sqrt{s}=7$~TeV.  This paper reports the result of a search for new phenomena in $\gamma+\missET$ events in $pp$ collisions at $\sqrt{s}=8$~TeV.

 The paper is organized as follows.
Section II gives a brief description of the ATLAS detector.
Section III explains the reconstruction of physics objects  and
Sec. IV describes the event selection applied.
Section V describes the  signal and background Monte Carlo simulation samples used.
Section VI outlines how the SM backgrounds are estimated and discusses the systematic uncertainties on the background estimation.
Section VII describes the results and their interpretation, and a summary is finally given in Sec. VIII.

\section{II. The ATLAS detector}
\label{sec:atlas}

The ATLAS detector \cite{ATLAS} is a multipurpose particle
physics apparatus with a forward-backward symmetric
cylindrical geometry and near 4$\pi$ coverage in solid angle~\cite{coord}.
The inner tracking detector (ID) covers the
pseudorapidity range $|\eta| <$ 2.5, and consists of a silicon
pixel detector, a silicon microstrip detector, and,
for $|\eta| <$ 2.0, a transition radiation tracker (TRT). The
ID is surrounded by a thin superconducting solenoid providing
a 2T magnetic field.
 A high-granularity lead/liquid-argon sampling electromagnetic
 calorimeter covers the region $|\eta|<$~3.2.
An iron/scintillator-tile calorimeter provides hadronic coverage in the range $|\eta| <$ 1.7.
The liquid-argon technology is also used for the hadronic calorimeters in the end-cap region 1.5~$<|\eta|<$~3.2 and for  electromagnetic and hadronic
measurements in the forward region up to $|\eta|=$~4.9.
The muon spectrometer (MS) surrounds
the calorimeters. It consists of three large air-core superconducting toroid systems, precision tracking chambers providing accurate muon tracking out to $|\eta|$~=~2.7, and additional detectors for triggering in the region $|\eta| <$~2.4.

\section{III. Event reconstruction  \label{Objects}}

Photons are reconstructed from clusters of energy deposits in the electromagnetic calorimeter measured in projective towers.
Clusters without matching tracks are classified as unconverted photon candidates.
A photon is considered as a converted photon candidate if it is matched to a pair of tracks that pass a TRT-hits requirement and form a vertex in the ID which is consistent with coming from a massless particle,
or if it is matched to a single track passing a TRT-hits requirement and having a first hit after the innermost layer of the pixel detector \cite{photonreco}.
The photon energy is corrected by applying the energy scales measured with
$Z \rightarrow e^+ e^-$ decays and cross-checked with $J/\psi \rightarrow e^+ e^-$ and $Z\rightarrow\ell\ell\gamma$ decays~\cite{egamma2014}.
Identification requirements are applied in order to reduce the contamination of the photon sample from $\pi^0$ or other neutral hadrons decaying to two photons.  The photon identification is based on the profile of the energy deposit in the first  and second layers of the electromagnetic calorimeter. Photons have to satisfy the tight identification criteria of Ref.~\cite{PhId}. They are also required to be isolated, i.e,  the energy in the calorimeters  in a cone of size $\Delta R = \sqrt{(\Delta \eta)^2 + (\Delta \phi)^2} = 0.4$ around the cluster barycenter,  excluding the energy associated with the photon cluster, is required to be less than $5\GeV$.
This cone energy is corrected for the leakage of the photon energy from the central core and for the
effects of multiple $pp$  interactions in the same or neighboring bunch crossings
superimposed on the hard physics process (referred to as  pileup interactions)  \cite{PhIsol}.

Electrons are reconstructed from clusters in the electromagnetic calorimeter matched to a track in the ID and
criteria for their identification and calibration procedure are similar to those used for photons.
Electron candidates must satisfy the medium++ identification requirement of Ref.~\cite{egamma2014}.
Muons are identified either as a combined track in the MS and ID systems,
or as an ID track that, once extrapolated to the MS, is associated with at least one track segment in the MS \cite{Aad:2014rra}.

Jets are reconstructed using the anti-$k_{t}$
algorithm~\cite{Cacciari:2008gp, Cacciari:2005hq} with a radius parameter $R = $ 0.4 from calibrated clusters of energy deposits in the calorimeters.  These clusters are seeded by calorimeter cells  with energies significantly above the measured noise.
The differences in calorimeter response between electrons, photons and hadrons are taken into account by classifying each cluster on the basis of its shape \cite{Aad:2011he},
prior to the jet reconstruction, as coming from an electromagnetic or hadronic shower.  The jet energy thus accounts for electromagnetic and hadronic energy deposits at the cluster level with correction factors derived from Monte Carlo (MC) simulation.
A further correction used to calibrate  the jet energy  to the scale of its constituent particles
\cite{Aad:2011he,JES2}, is then applied.
Jets are required to have transverse momentum \pt\ $>$ 30~\GeV, $|\eta| < 4.5$, and a distance to the closest preselected electron or photon of $\Delta R>0.2$.

The vector momentum imbalance in the transverse plane is obtained
from  the negative vector sum of the reconstructed and calibrated physics objects
and  is referred to as missing transverse momentum, \metv.
The symbol $\met$ is used for its magnitude.
Calorimeter energy deposits are associated with a reconstructed and identified high-$\pt$ object in a specific order: electrons with \pt\ $>10\GeV$, photons with \pt\ $>10\GeV$, and  jets with \pt\ $>20\GeV$.
Deposits not associated with any such objects are also taken into account in the \metv\ calculation \cite{Report No. ATLAS-CONF-2013-082} using an energy-flow algorithm that considers calorimeter energy deposits as well as ID tracks \cite{Aad:2012re}.

\section{IV. Event selection \label{Selection}}

The data were collected in $pp$ collisions at $\sqrt{s}=8$ \TeV.
Events were selected using an \met\ trigger that requires a missing transverse momentum greater than 80 GeV \cite{MET_Trig}.
Events selected using an $e/\gamma$ trigger with a threshold of $p_{\textrm{T}} > 120\GeV$ are also used in some control regions as described below \cite{Egamma_Trig}.
Only data taken during periods when the calorimeters, ID and MS  were well functioning are considered.
The data used correspond to an integrated luminosity  of 20.3 \ifb.
Quality requirements are applied to photon candidates in order to reject those arising from instrumental problems.
In addition, quality requirements are applied in order to remove jets arising from detector noise and out-of-time energy deposits in the calorimeter from cosmic rays or other noncollision sources \cite{jetcleaning2}.
Events are required to have a reconstructed primary vertex with at least five associated tracks;
the primary vertex is defined as  the vertex with the highest sum of the squared transverse momenta of its
associated tracks.

The criteria for selecting events in the signal region (SR) are optimized to have good acceptance for the squark model  and the 
dark--matter model with a $Z'$-like mediator, described in Sec. I, as well as to suppress the background from SM processes. This signal region also provides good sensitivity to the other models described in Sec. I.
Events in the SR are required to have $\MET > 150$~\GeV and
a photon with  $\pt^{\gamma}>125 \GeV$  and $|\eta|<1.37$.
It is also required that the photon and \metv\ are not overlapped in azimuth: $\Delta\phi(\gamma, $\metv$) > 0.4$.
Events with more than one jet or with a jet with $\Delta\phi(\jet,$\metv$)<0.4$ are rejected.  Events with one jet are retained to increase the signal acceptance and reduce systematic uncertainties related to the modeling of initial-state radiation.
 Events are required to have no electron  ($\pt>7\GeV$, $|\eta| < 2.47$) and no muon ($\pt>6\GeV$, $|\eta| < 2.5$).
The lepton veto mainly rejects $W/Z$ events with charged leptons in the final state.
For events satisfying these criteria, the $\MET$ trigger efficiency is $0.99 \pm 0.01$, as determined using events selected with the $e/\gamma$ trigger.
The final data sample contains 521 events, where 319 and 202 events have zero and one jet, respectively.

\section{V. Monte Carlo simulation samples \label{simsample}}

Monte Carlo simulated samples are used to estimate the signal acceptance, the detector efficiency and to help in the estimation of the SM background contributions.

Simulated signal samples for ADD models are generated with $\pythiaeight$ \cite{Sjostrand:2007gs} version 1.7.5 using the MSTW2008LO~\cite{Martin:2009iq} parton distribution function (PDF) set.
Simulations were run for two values ($2.0$ and $2.5\TeV$)
of the scale parameter $M_{\rm D}$ and with the number of extra dimensions, $n$, varied from two to six.

Simulated samples of dark--matter production $pp\rightarrow\gamma+\chi\bar{\chi} + X$ via the $qq\chi\bar{\chi}$ interaction are generated using $\madgraph5$~\cite{Alwall:2011uj}  version 1.4.8.4,
with showering and hadronization modeled by $\pythiaeight$ version 1.6.5 using the set of parameters optimized to describe the properties of the events referred to as AU2 tune~\cite{Report No. ATL-PHYS-PUB-2011-008}; the MSTW2008LO PDFs are used. Values of $m_\chi$ from 1 to 1300 GeV are considered.
In addition, simulated samples of $pp\rightarrow\gamma+\chi\bar{\chi}$ are produced using the simplified model with a $Z'$-like mediator \cite{LMWIMP} using the same simulation programs as for the EFT samples.
Vector and axial-vector couplings are both considered. For each value of the mediator mass $m_V$, two different values of the mediator width are simulated: $\Gamma=m_V/8\pi$ and $\Gamma=m_V/3$.
The smaller value corresponds to a mediator that can annihilate into only one quark flavor and helicity and has unit couplings; it can be regarded as an approximate lower limit on the mediator width.
A value of $\Gamma= m_V/3$ is a reasonable upper bound for a narrow resonance approximation.

Samples of $pp\rightarrow\gamma+\chi\bar{\chi} + X$ are also produced via the $\gamma\gamma\chi\bar{\chi}$ interaction model~\cite{schannelWIMP} with a fermionic WIMP. These samples are generated with $\madgraph$5 version 1.4.2 for
a WIMP mass of 130 GeV and over a grid of values of $k_1$ and $k_2$.

Simulated samples of $pp\rightarrow\tilde{q}\tilde{q}^\ast\gamma + X \rightarrow q\bar{q} \gamma+\tilde{\chi}_1^0\tilde{\chi}_1^0 + X$ are generated with $\madgraph$5 version 1.5.11 with showering and hadronization modeled by $\pythia~6$ \cite{PYTHIA} version 4.2.7 and CTEQ6L1 PDFs \cite{Pumplin:2002vw}, with the requirement of having one photon at parton level  with $\pt^\gamma >80\GeV$ and $|\eta|<2.5$.  Only the first two generations of squarks are considered, and they are assumed to be degenerate in mass.
Signal cross sections are calculated to next-to-leading order in the strong coupling constant including the resummation of  soft gluon emission at next-to-leading-logarithm accuracy when available~\cite{Beenakker:1996ch,Kulesza:2008jb,Kulesza:2009kq,Beenakker:2009ha,Beenakker:2011fu}.
The nominal cross section and its uncertainty are taken from an envelope of cross-section predictions using different PDF sets and factorization and renormalization scales, as described in Ref.~\cite{Kramer:2012bx}.

Simulated samples of $Z\gamma$ and $W\gamma$ events  are generated with $\sherpa$ version 1.4.1 \cite{Gleisberg:2008ta}, with parton-level requirements of $\pt^\gamma>$ 70 GeV  and $\pt^\gamma>$ 80 GeV, respectively, and dilepton invariant  mass $m_{\ell\ell} > 40 \GeV$.
A sample of simulated $\gamma+\jet$ events is generated with $\pythiaeight$ version 1.6.5.
The $W/Z+\jet$ processes are also simulated using $\sherpa$ version 1.4.1 with massive $b/c$ quarks.
Diboson samples are generated with $\herwig$  \cite{Corcella:2000bw, Corcella2} version 6.520, the single-top samples with $\mcatnlo$ \cite{Frixione:2002ik, Frixione2} version 4.06 for $s$-channel and $Wt$ production, and $\acermc$ \cite{Kersevan:2004yg} version 3.8 for $t$-channel production. Simulated samples of top-quark pair production are generated with $\powheg$~\cite{powheg1} version r2129.

$\herwig$ version 6.520 is used for simulating the parton shower and fragmentation processes in combination with $\jimmy$ \cite{Butterworth:1996zw} for underlying-event simulation for the $\mcatnlo$  samples, while $\pythia~6$ version 4.2.6 is used for the $\powheg$ and $\acermc$  samples.
 The proton PDFs used are CTEQ6L1 \cite{Pumplin:2002vw} for the $\pythiaeight$ and $\acermc$ samples, and CT10 \cite{Lai:2010vv}
for the $\mcatnlo$, $\sherpa$,  and $\powheg$ samples.
The ATLAS underlying-event tune AUET2  \cite{Report No. ATL-PHYS-PUB-2011-008} is used, except for the $t\bar{t}$ sample, which uses
the new Perugia 2011C tune \cite{pythiaperugia}. $\sherpa$ uses its own parton shower, fragmentation and underlying-event model.

Differing pileup conditions as a function of the instantaneous luminosity are taken into
account by overlaying simulated
minimum-bias events generated with   $\pythiaeight$ onto the hard-scattering process and reweighting
their number according to the observed distribution of the average number of interactions per beam crossing.

The simulated samples are processed either with a full ATLAS detector
simulation~\cite{:2010wqa} based on
GEANT4~\cite{Agostinelli:2002hh} or a fast simulation based on the
parametrization of the response to the electromagnetic and hadronic
showers in the ATLAS calorimeters~\cite{atlfast} and a simulation of
the trigger system.
The results based on fast simulation are validated against fully simulated
samples.
The simulated events are
reconstructed and analyzed with the same analysis chain as for the data,
using the same trigger and event
selection criteria discussed in Sec. IV.

\section{VI. Background estimation \label{BGEstimation}}

The SM background to the $\gamma+\missET$ final state
 is dominated by the $Z(\rightarrow \nu \nu) + \gamma$ process, where the photon is due to initial-state radiation.
Secondary contributions come from $W\gamma$  and $Z\gamma$ production with unidentified electrons, muons or hadronically decaying $\tau$ leptons, or $W/Z$ production where a lepton or an associated radiated jet is misidentified as a photon. In addition, there are smaller contributions from top-quark pair, diboson, $\gamma$+jet and multijet production.

\subsection{A. $Z\gamma$ and $W\gamma$ backgrounds}
The $\missET$ distribution of events due to $Z\gamma$ and $W\gamma$ backgrounds is described using simulated samples, while the normalization is obtained via a likelihood fit to observed yields in several control regions (CRs), constructed to be enriched in specific backgrounds.
  Poisson likelihood functions are used for event counts in all regions; the systematic uncertainties described in Sec. VI E are treated as Gaussian-distributed nuisance parameters in the likelihood function.
  Key ingredients of the fit are the normalization scale factors for the $W\gamma$ and $Z\gamma$ processes, which  enable  observations in  the CRs to constrain background estimates in the SR.  The same normalization factor is used for $Z(\nu\nu)+\gamma$, $Z(\mu\mu)+\gamma$, and $Z(ee)+\gamma$ events.

  Three control regions are defined by inverting lepton vetoes.
 In the first control region, the $W\gamma$ contribution is enhanced by requiring the presence of a muon. The second (third) control
 region enhances the $Z\gamma$ background by requiring the presence of a pair of muons (electrons).
   In the muon control region,  in order to ensure that the $\missET$ spectrum is similar
to the one in the signal region, muons are treated as invisible particles in the $\missET$ calculation.
The same procedure is followed for electrons in the electron control region.
  In each case, the CR lepton selection follows the same requirements as the SR lepton veto with the additional requirements that the lepton must be associated with an ID isolated track and that $\Delta R(\ell,\gamma)>0.5$.  In addition, the photon pseudorapidity requirement is relaxed with respect to the SR selection: $|\eta| < 2.37$, excluding the calorimeter barrel/end-cap transition region $1.37<|\eta|<1.52$, to increase the number of events in the CR. In both the $Z\gamma$-enriched control regions, the dilepton mass $m_{\ell \ell}$ is required to be greater than $50 \GeV$.  The normalization of the dominant $Z\gamma$ background process is largely constrained by the event yields in the two $Z(\ell\ell)+\gamma$ control regions.
 The results are cross-checked using the transfer-factor technique employed in the previous ATLAS analysis of the $\gamma+\missET$ final state~\cite{Aad:2012fw}; the two methods give consistent results.

\subsection{B. Fake photons from misidentified electrons}
Contributions from processes in which an electron is misidentified as a photon are estimated by scaling  yields from a sample of $e +\missET$ events by an electron-to-photon misidentification factor. This factor is measured in mutually exclusive samples of $e^+e^-$ and $\gamma+e$ events. To establish a pure sample of electrons,  $m_{ee}$ and $m_{e\gamma}$ are both required to be consistent with the $Z$ boson mass, and the multijet background estimated from sidebands is subtracted.
The  misidentification factor is parametrized as a function of $\pt$  in three pseudorapidity bins.   Similar estimates are made for the three control regions with leptons, scaling event yields from samples matching the
      control region requirements, but requiring an electron
      rather than a photon.

\subsection{C. Fake photons from misidentified jets}
Background contributions from events in which a jet is misidentified as a photon are estimated from samples of $\gamma+\missET$ events where the photon does not fulfill  the isolation requirement. The yield in this sample is scaled by a jet-to-photon misidentification factor,  after subtraction of the contribution from real photons.
The jet-to-photon misidentification factor is measured in samples enriched in jets, selected by inverting some photon identification criteria, and is determined from the ratio  of  isolated jets to nonisolated jets.
This estimate also accounts for the contribution from multijets, which can mimic the monophoton signature if one jet is misreconstructed as a photon and one or more of the other jets are poorly reconstructed, resulting in large fake $\MET$.
The multijet  background is found to be negligible in the SR.

\subsection{ D. \gammajets\ background}
The \gammajets\ background in the signal region consists of events where the jet is poorly
reconstructed and partially lost, creating  fake $\MET$.
Despite the large production rate, this process is only a minor source of background as it is suppressed by the large $\MET$ and the large jet--$\missET$ separation requirements in the SR.
This background is estimated from MC simulation and is cross-checked with a data-driven estimate,
which gives a result in agreement with the MC simulation, but is limited by a large statistical uncertainty. The data-driven estimate is derived from  a control region defined by requiring all the selection criteria of the SR but reversing the $\Delta\phi(\jet,$\metv$)$ requirement, thereby selecting poorly reconstructed events in which the jet is aligned with the \metv.   Simulated samples are used to estimate and subtract electroweak backgrounds coming from \wzjet~and $Z/W+\gamma$ processes. As events with a jet with $\pt>$ 30 \GeV\ that is not well separated from \metv\ are vetoed in the SR selection, the  \gammajets\ and multijet contribution in the SR is then estimated with a linear extrapolation of the jet \pt\ spectrum in this CR to the $\pt < 30$ GeV region.

\subsection{E. Final estimation and systematic uncertainties}
Background estimates in the SR are first derived from a fit using only data from the lepton CRs,
in order to assess whether the observed SR yield is consistent with the background model.
The values of the normalization factors for the $W\gamma$ and $Z\gamma$ backgrounds obtained from the fit to the CRs are $k_{W\gamma}=0.81\pm0.05 (\rm stat.) \pm0.06 (\rm syst.)$ and $k_{Z\gamma}=0.89\pm0.08 (\rm stat.) \pm0.08 (\rm syst.)$,
where the systematic error takes into account the various sources of systematic uncertainties described below.  Distributions of the missing transverse momentum in the three control regions are shown in Figs.~\ref{fig:auxiliary_1mu}--\ref{fig:auxiliary_2e}.

The techniques used for the background estimation are checked in a validation region, where events are selected with the same criteria as used for the signal region, except for a lower $\met$ ($110$--$150\GeV$) and a larger photon pseudorapidity  range ($\abseta<2.37$, excluding the calorimeter barrel/end-cap transition region $1.37<|\eta|<1.52$) to increase the statistical power.
To suppress the background from $\gammajets$ events and from fake photons to a level similar to that in the SR, a requirement on the azimuthal separation between the photon and the jet -- when there is a jet in the event -- is applied: $\Delta\phi(\gamma,$ jet$) < 2.7$.
To minimize the contamination of this region by signal events, a requirement on the azimuthal separation between the photon and \metv\ is  added:  $\Delta\phi(\gamma, $\metv$) < 3.0$.
The number of events in the data in this region is $307$ and the estimated total background,
obtained from the background-only fit to the control regions,
 is $272 \pm 17 \pm 14$, resulting in  agreement between the data and expectation within  $ 2 \sigma$. Detailed results are shown in Table~\ref{tab:SR_yields}; systematic uncertainties are computed as described below for the SR.

Systematic uncertainties on the background predictions in the signal region are presented here as percentages of the total
background prediction.
This prediction is obtained  from the CR fit, which provides constraints on many of the sources of systematic uncertainty.  The dominant contribution is due to the uncertainty on the electron fake rate, which contributes a 4.6\% relative uncertainty, and to the reconstruction and identification efficiency corrections applied to electrons and muons in MC simulation, which contribute 1.3\% and 0.7\% relative uncertainty, respectively.
 The uncertainty on the absolute electron/photon energy scale translates into a 0.6\% relative uncertainty on the total background prediction.      Uncertainties in the simulation of the electron/photon energy resolution, isolation, and identification efficiency contribute a relative uncertainty
of 0.1\% on the total predicted background.
 The uncertainty on the absolute jet energy scale~\cite{JES2} and the jet energy resolution \cite{JER} contribute 0.1\%
and 0.5\% relative uncertainties, respectively.
 Uncertainties on the scale and resolution of the calorimeter energy deposits not associated with high-$\pt$ physics objects affect the calculation of the \metv\ and generate an uncertainty of 0.3\% on the background prediction.
 Uncertainties on the PDF are evaluated by following, for the CT10
  and MSTW2008LO PDF sets, the PDF4LHC recommendations \cite{Botje01}.
 The Hessian method is used to obtain asymmetric uncertainties at $68\%$ confidence level (C.L.).
In addition, to obtain inter-PDF uncertainties, the results are then compared with those obtained with the NNPDF set.
Renormalization and factorization scale uncertainties are also taken into account by increasing and decreasing the scales used in the MC generators by a factor of 2. PDF and scale uncertainties contribute 0.7\% to the background prediction uncertainty.
 After the fit, the uncertainty on the jet energy scale due to corrections for
pileup, and the uncertainties on the trigger efficiency and luminosity \cite{lumi2013}, are
found to have a negligible impact on the background estimation.
The  final  total background prediction systematic uncertainty is about 5\%, while the statistical uncertainty is about 6\%.

\section{VII. Results}

Table \ref{tab:SR_yields} presents the observed number of events and the SM background predictions obtained from a fit to the CRs. The $\missET$ distribution in the SR is shown in Fig.~\ref{fig:SR_event}.

\begin{figure}\centering
\includegraphics[width=1.0\linewidth]{{{fig_06}}}
\caption{Distribution of $\missET$ in the data and for the expected background in the single-muon control region. The total background expectation is normalized to the observed number of events in this control region.
The dashed band includes statistical and systematic uncertainties.
Overflows are included in the final bin. The lower part of the figure shows
the ratios of data to expected-background event yields.
}\label{fig:auxiliary_1mu}
\end{figure}

\begin{figure}\centering
\includegraphics[width=1.0\linewidth]{{{fig_07}}}
\caption{Distribution of $\missET$ in the data and for the expected background in the two-muon control region. The total background expectation is normalized to the observed number of events in this control region.
The dashed band includes statistical and systematic uncertainties.
 Overflows are included in the final bin. The lower part of the figure shows
 the ratios of data to expected-background event yields.}\label{fig:auxiliary_2mu}
\end{figure}

\begin{figure}\centering
\includegraphics[width=1.0\linewidth]{{{fig_08}}}
\caption{Distribution of $\missET$ in the data and for the expected background in the two-electron control region. The total background expectation is normalized to the observed number of events in this control region.
The dashed band includes statistical and systematic uncertainties.
Overflows are included in the final bin. The lower part of the figure shows
the ratios of data to expected-background event yields.}\label{fig:auxiliary_2e}
\end{figure}

\begin{table}[!htbp]
\centering
\caption{Observed event yield compared to  predicted event yield  from SM backgrounds in the SR and the validation region (VR), using estimates and uncertainties obtained from a fit in the control regions.  Uncertainties are statistical followed by systematic. In the case of the $\gammajets$ process a global uncertainty is quoted.
}
\label{tab:SR_yields}

\begin{tabular}{lrr}
\hline\hline
Process & Event yield (SR) & Event yield (VR) \\ \hline
$\gammaznunu$                     & $389\pm36 \pm 10$  & $153 \pm 16 \pm 10$\\
$\wgamma$                         & $ 82.5\pm 5.3 \pm 3.4$ & $67 \pm 5 \pm 5$  \\
$\wzjet, t\bar{t},\text{diboson}$ & $83 \pm 2 \pm 28$ & $47 \pm 2 \pm 14$ \\
$\zgamma$                         & $2.0\pm 0.2 \pm 0.6$ & $2.9 \pm 0.3 \pm 0.6$\\
$\gammajets$                      & $0.4 _{-0.4}^{+0.3} $ & $2.5 _{-2.5}^{+4.0}$\\
\hline
Total background & $557\pm36 \pm 27$ & $272 \pm 17 \pm 14$\\
Data & $  521 $ & 307 \\
\hline\hline
\end{tabular}
\end{table}

\begin{figure}[!htbp]
\centering
\includegraphics[width=1.0\linewidth]{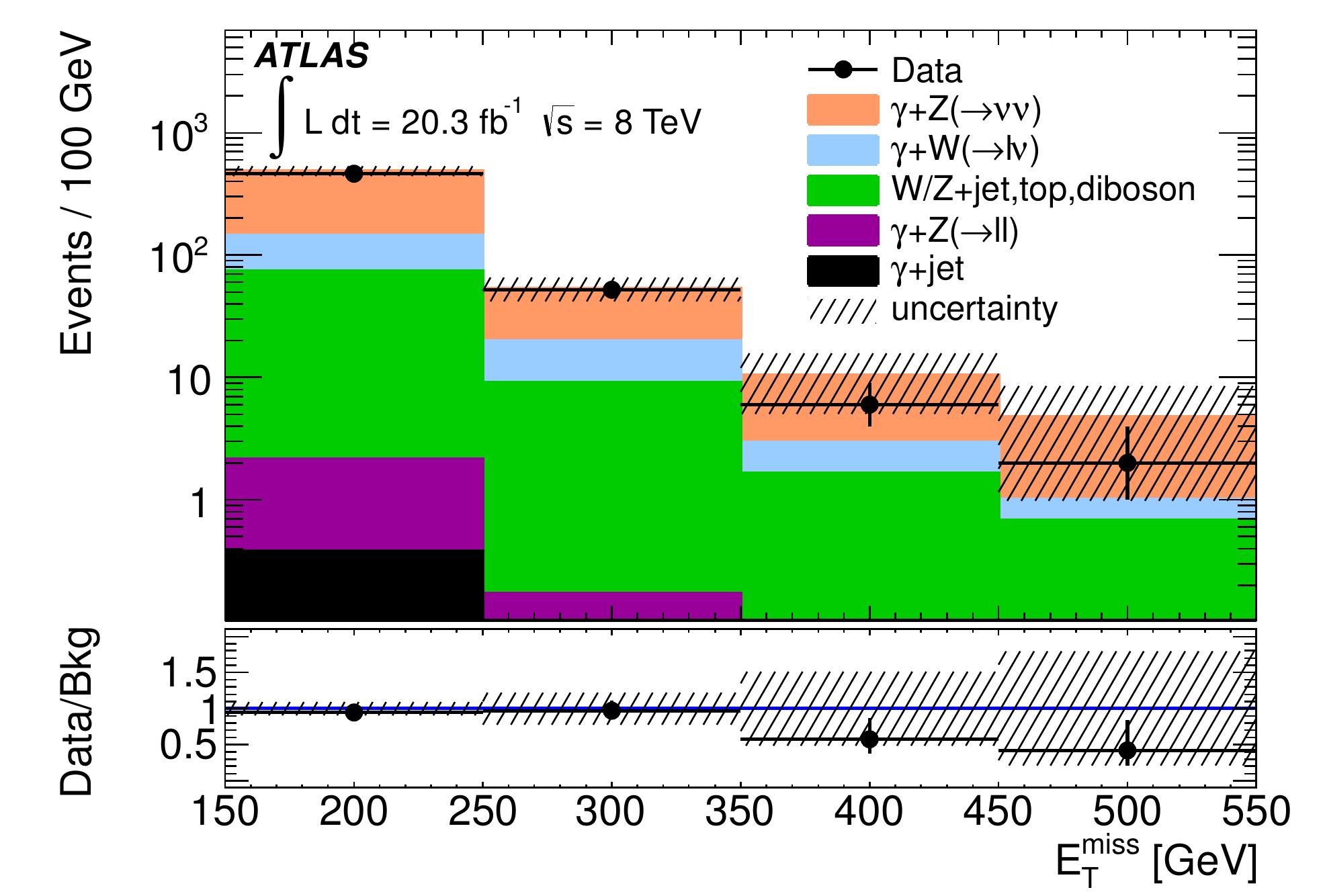}
\caption{Distribution of $\missET$ in the signal region for data and for the background predicted from the fit in the CRs.
The dashed band includes statistical and systematic uncertainties.
Overflows are included in the final bin.
The lower part of the figure shows
the ratios of data to expected-background event yields.}
\label{fig:SR_event}
\end{figure}

As the $521$ events observed in data are well described by the SM background prediction of $557 \pm 36 \pm 27$, the results are interpreted in terms of exclusions on models that would produce an excess of  $\gamma+\missET$ events. Upper bounds are calculated using a one-sided profile likelihood ratio and the $CL_S$ technique~\cite{Read:2002hq,Junk:1999kv}, evaluated using the asymptotic approximation~\cite{Cowan:2010js}, making use of data in the CRs as well as in the SR.

The most model-independent limits provided are those on the fiducial cross section of a potential new physics process, $\sigma\times A$, where $\sigma$ is the cross section and $A$ is the fiducial acceptance. The latter is defined using a selection identical to that defining the signal region but applied at particle level, where the particle-level \metv\ is the vector sum of invisible particle momenta.   The limit on $\sigma\times A$ is derived from a limit on the visible cross section $\sigma\times A \times \epsilon$, where $\epsilon$ is the fiducial reconstruction efficiency. A conservative estimate $\epsilon = 69$\% is computed  using  ADD and WIMP samples with no quark/gluon produced from the main interaction vertex. The expected (observed) upper limit on the fiducial cross section is 6.1 (5.3) fb at 95\% C.L. and 5.1 (4.4) fb at 90\% C.L.  These limits are applicable to any model that produces $\gamma+\missET$ events in the fiducial region and has similar reconstruction efficiency $\epsilon$.

For limits on specific models, the impact of systematic uncertainties on signal samples is evaluated separately for $A\times\epsilon$  [PDF, scale, initial-state radiation (ISR), and final-state radiation (FSR) uncertainties] and the cross section $\sigma$ (PDF and scale uncertainties). Only uncertainties affecting $A\times\epsilon$ are included in the statistical analysis; uncertainties affecting the cross section are indicated as bands on observed limits and written as $\sigma_{\textrm{theo}}$.
For the EFT and simplified-model DM samples, scale uncertainties are evaluated by
varying the renormalization, factorization and matching scales in $\madgraph$  by a factor of 2.   For the ADD samples, the \pythiaeight~renormalization and factorization scale parameters are varied independently to $0.5$ and $2.0$. For these samples, the ISR and FSR signal uncertainties are assessed by varying the \pythiaeight parameters, as done in Ref.~\cite{Aad:2014vka}.  For the squark model described in Sec.~I, systematic uncertainties arising from the treatment of ISR/FSR are studied
with MC event samples by varying the value of $\alpha_{\rm s}$; the renormalization and factorization scales and the
$\madgraph/\pythia$ matching parameter are also varied to estimate the related uncertainties.
Radiation uncertainties are typically less than 10\%, PDF uncertainties less than 30\%, and scale uncertainties less than 20\%.

Limits on dark--matter production are derived from the cross-section limits at a given WIMP mass $m_\chi$, and expressed as 90\% C.L. limits on the suppression scale $\Mstar$, for the D5 (Fig.~\ref{fig:Plot_D5_Limits}), D8 (Fig.~\ref{fig:Plot_D8_Limits}),
and D9 (Fig.~\ref{fig:Plot_D9_Limits}) operators.
Values of $\Mstar$ up to 760, 760, and 1010 GeV are excluded for the D5, D8, and D9 operators, respectively.
As already mentioned, the  effective field theory model becomes a poor approximation when the momentum transferred in the interaction, $Q_{\textrm{tr}}$, is comparable to the mass of the intermediate state $m_V=\Mstar\sqrt{g_fg_\chi}$~\cite{Goodman:2010ku,Busoni:2013lha}.
In order to illustrate the sensitivity to the unknown ultraviolet completion of the theory, limits computed retaining only simulated events with $Q_{\textrm{tr}}<m_V$ are also shown, for a value of the coupling $\sqrt{g_fg_\chi}$  equal to either unity or the maximum value ($4\pi$) that allows the perturbative approach to be valid. This procedure is referred to as truncation.
As can be seen in  Figs.~\ref{fig:Plot_D5_Limits}--\ref{fig:Plot_D9_Limits}, the truncated limits nearly overlap with the nontruncated limits for a $4\pi$ coupling.
For unit coupling, the truncated  limits are less stringent than the nontruncated limits at low $m_{\chi}$, and the analysis loses sensitivity for \mbox{$m_{\chi} >100\ (200)$\ \GeV}\ for the D5 and D8 (D9) operators.
In this case, for the D5 and D8 operators, as no sample was generated between $m_\chi=50$ GeV and $m_\chi=100$ GeV, the limit is only shown up to $m_\chi=50$ GeV; for the D9 operator, as no sample was generated between $m_\chi=100$ GeV and $m_\chi=200$ GeV, the limit is only shown up to $m_\chi=100$ GeV.
These lower limits on $\Mstar$ can be translated into upper limits on the WIMP--nucleon interaction cross section as a function of $m_\chi$ using Eqs.~(4) and (5) of Ref.~\cite{Goodman:2010ku}. Results are shown in Fig.~\ref{fig:Plot_D59_Limits} for spin-independent (D5) and spin-dependent (D8, D9) $\chi$--nucleon interactions and are compared to measurements from various DM search experiments~\cite{xenon100,cdms,cogent,picasso,simple,LUX,Desai:2004pq,icecube,coupp2012,Angloher:2011uu,Agnese:2013rvf,Bernabei:2008yi,Abramowski:2013ax}.
The search for dark--matter pair production in association with a $\gamma$ at the LHC extends the limits  on the $\chi$--nucleon scattering cross section into the low-mass region $m_{\chi}<10\GeV$ where the astroparticle experiments have less sensitivity due to the very low-energy recoils such low-mass DM particles would induce.

\begin{figure}[!htbp]
\centering
\includegraphics[width=1.0\linewidth]{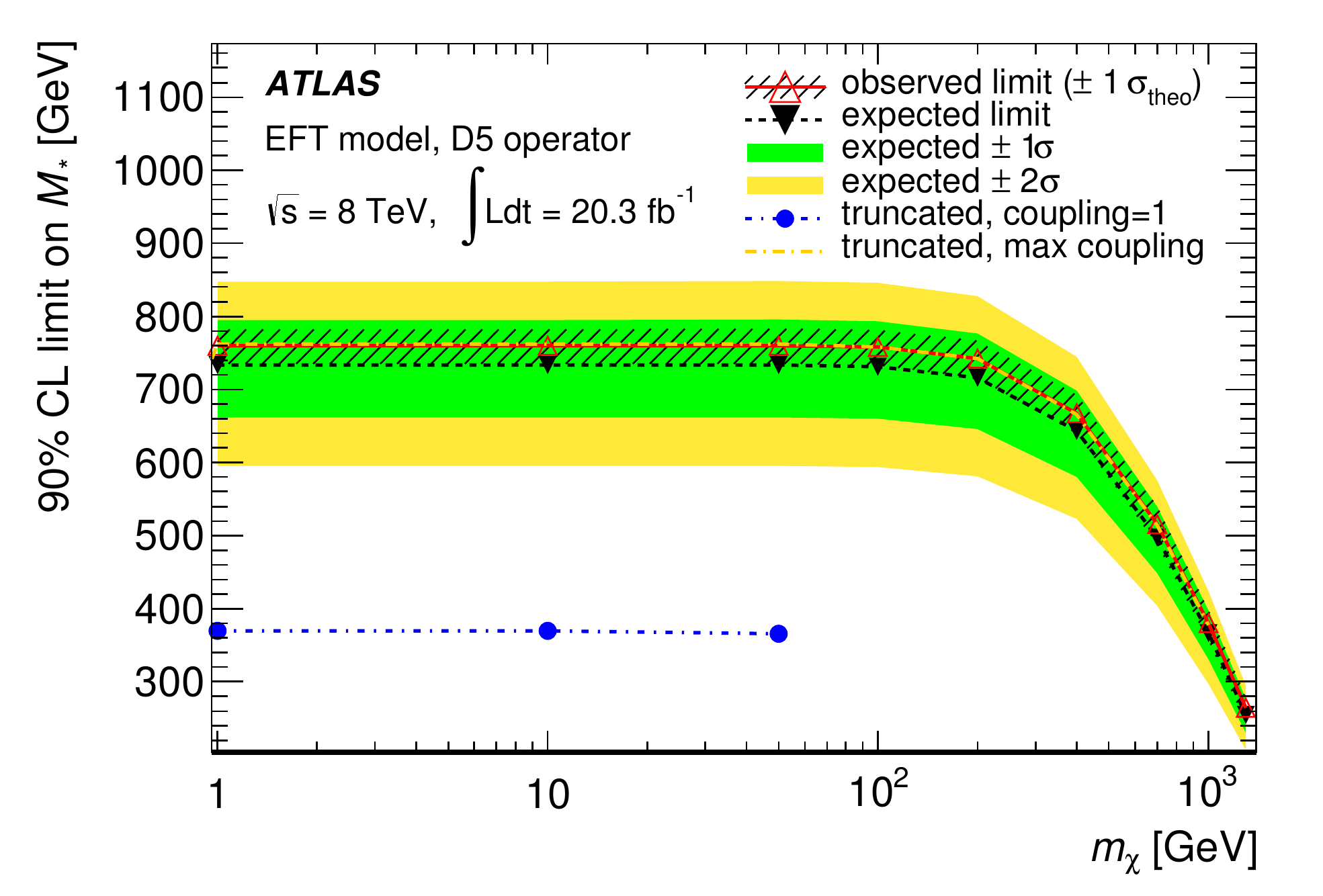}
\caption{Limits at 90\%  C.L. on the EFT suppression scale $\Mstar$ as a function of the WIMP mass $m_{\chi}$,
    for the vector operator D5. Results where EFT truncation is applied (see text) are also shown, assuming coupling values
    $\sqrt{g_fg_\chi}=1,4\pi$.
 }
\label{fig:Plot_D5_Limits}
\end{figure}

\begin{figure}[!htbp]
\centering
\includegraphics[width=1.0\linewidth]{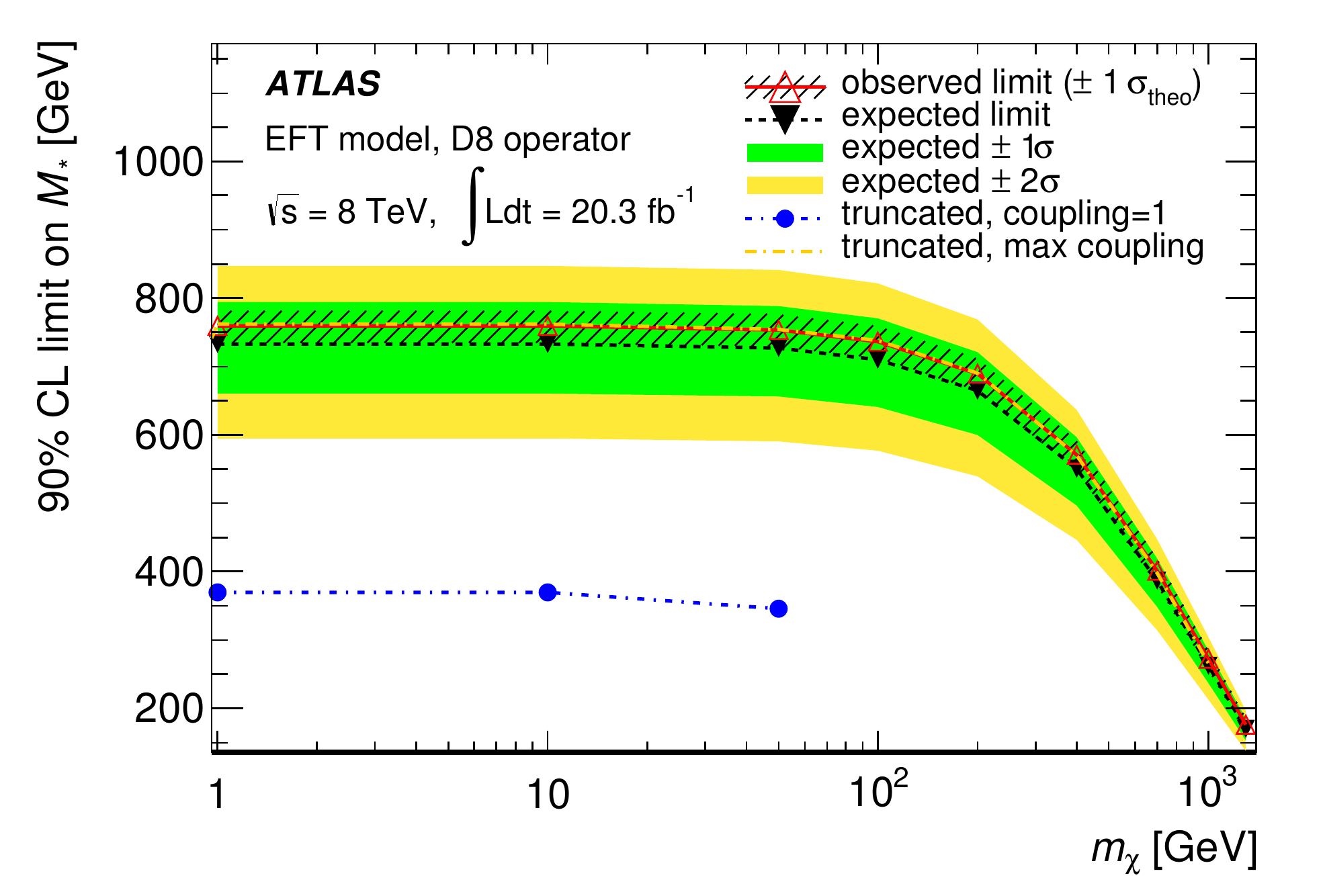}
\caption{Limits at 90\%  C.L. on the EFT suppression scale $\Mstar$ as a function of the WIMP mass $m_{\chi}$,
    for the axial-vector operator D8. Results where EFT truncation has been applied (see text) are also shown, assuming coupling values $\sqrt{g_fg_\chi}=1,4\pi$.
    }
\label{fig:Plot_D8_Limits}
\end{figure}

\begin{figure}[!htbp]
\centering
\includegraphics[width=1.0\linewidth]{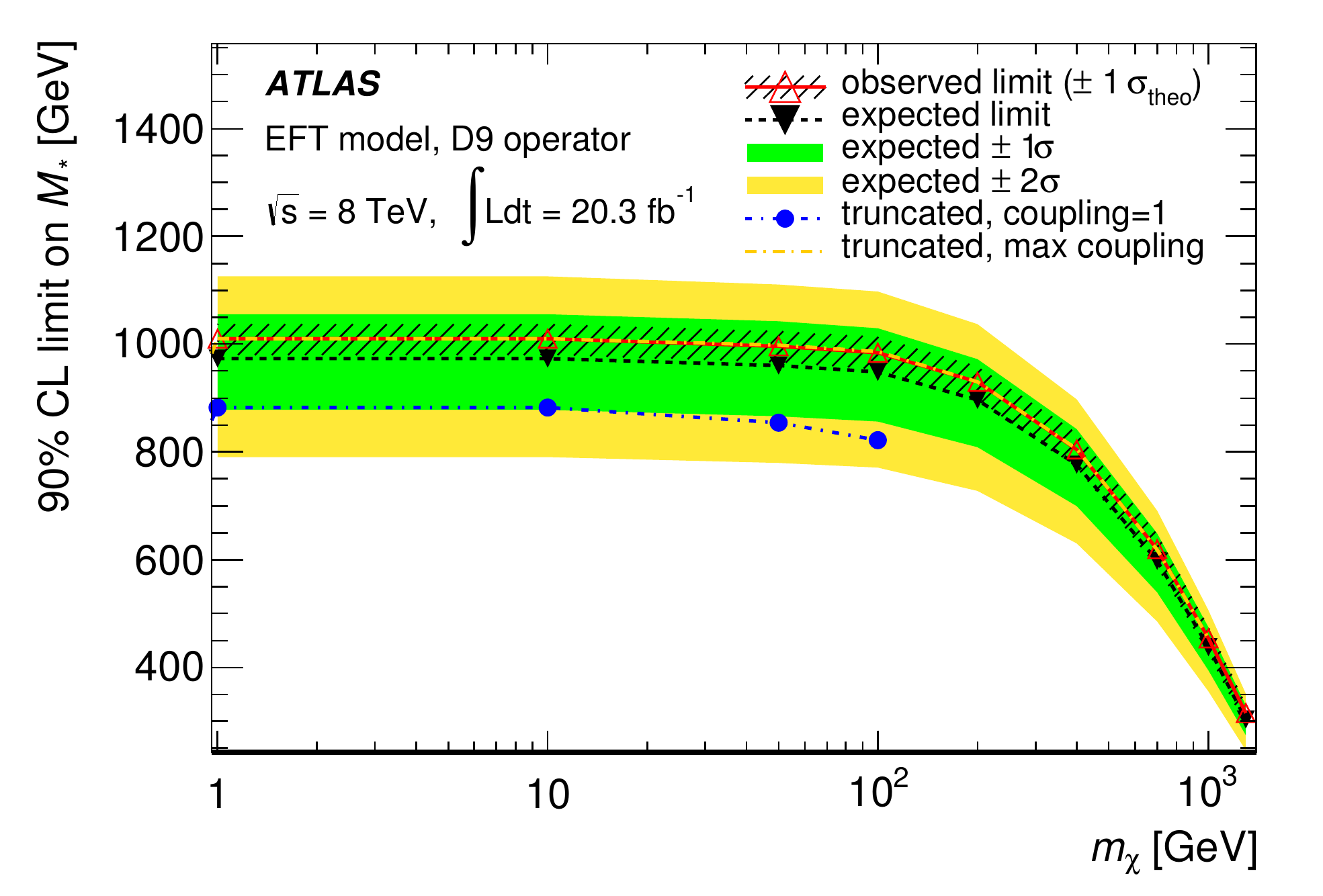}
\caption{Limits at 90\%  C.L. on the EFT suppression scale $\Mstar$ as a function of the WIMP mass $m_{\chi}$,
    for the tensor operator D9. Results where EFT truncation is applied (see text) are also shown, assuming coupling values
    $\sqrt{g_fg_\chi}=1,4\pi$.
    }
\label{fig:Plot_D9_Limits}
\end{figure}

\begin{figure*}[!htbp]
\centering
\includegraphics[width=0.7\linewidth]{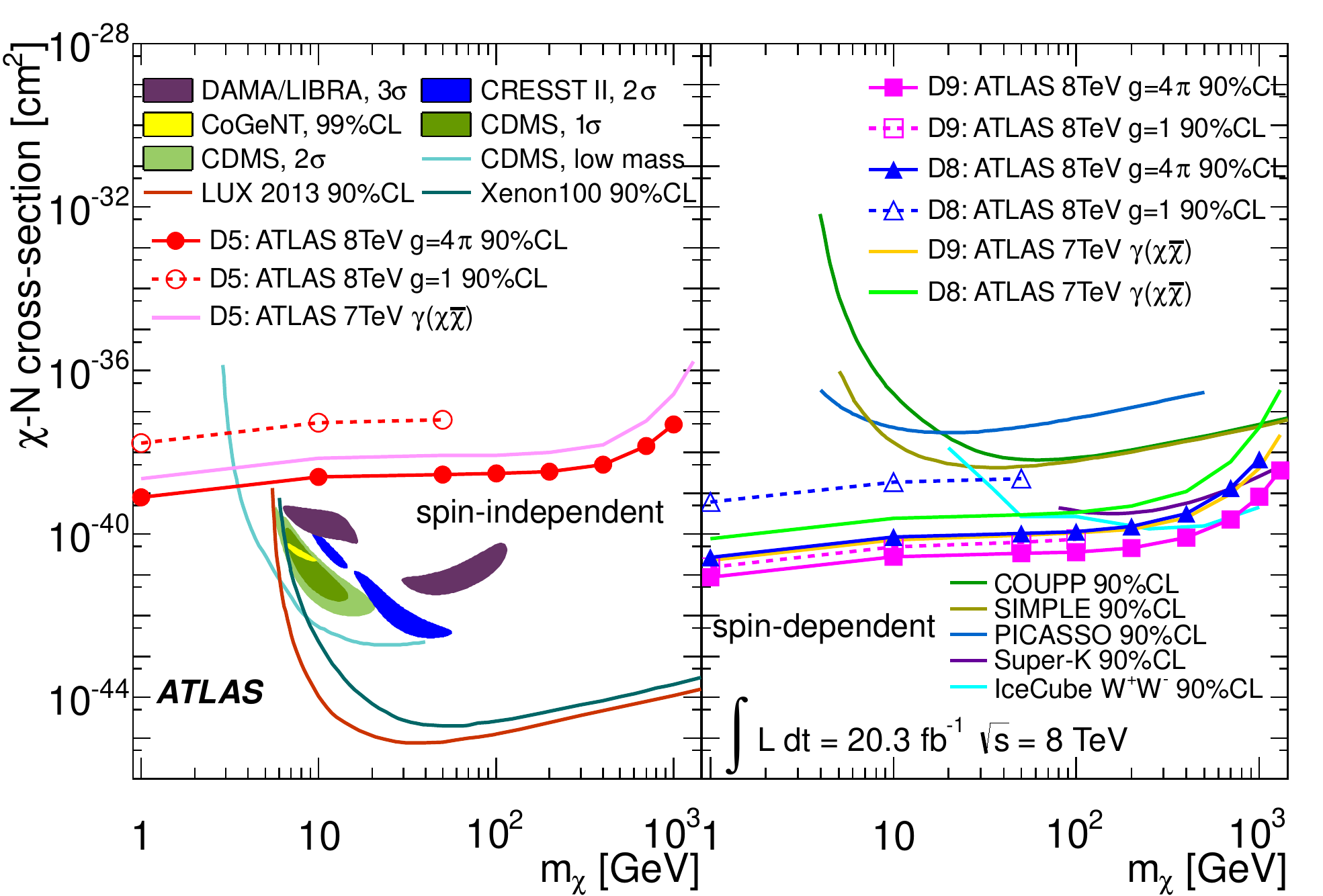}\label{VH_2}
\caption{  Upper limits at 90\% C.L. on the WIMP--nucleon ($\chi$-N) scattering cross section as a function of $m_{\chi}$
for spin-independent (left) and spin-dependent (right) interactions,
for a coupling strength $g=\sqrt{g_fg_\chi}$ of unity or the maximum value ($4\pi$) that keeps the model within its perturbative regime.
  The truncation procedure is applied for both cases.
 The results obtained from ATLAS with $7\TeV$ data for the same channel are shown for comparison.
 Also shown are results from various dark--matter search experiments~\cite{xenon100,cdms,cogent,picasso,simple,LUX,Desai:2004pq,icecube,coupp2012,Angloher:2011uu,Agnese:2013rvf,Bernabei:2008yi,Abramowski:2013ax}.}
\label{fig:Plot_D59_Limits}
\end{figure*}

Simplified models with explicit mediators are ultraviolet complete and therefore robust for all values of $Q_{\textrm{tr}}$.   For the simplified $Z'$-like model with vector interactions and  mediator width $\Gamma=m_V/3$, Fig.~\ref{fig:Plot_VA_Limits} shows the 95\% C.L. limits on the coupling parameter $\sqrt{g_fg_\chi}$ calculated for various values of the WIMP and mediator particle masses, and compared to the lower limit resulting from the relic DM abundance \cite{WMAP}.
  In the region above the dashed line, the lower limits on the coupling resulting from
  the relic abundance of DM are higher than the upper limits found in this analysis.
Figures~\ref{fig:lightM_Lambda_DM50_400_A} and \ref{fig:lightM_Axial_Lambda_DM50_400_A} show, for vector and axial-vector interactions and different values of the WIMP mass, the corresponding 95\% C.L. limits on the suppression scale $\Mstar$ as a function of $m_V$. One can note how, when the mediator mass is greater than the LHC reach, the EFT model provides a good approximation of the simplified model with $\Mstar= m_V/\sqrt{g_f g_{\chi}}$. The truncation procedure is applied when computing the EFT limits; these limits are always more conservative than those from the simplified model as long as $m_V$ is greater than or equal to the value used for EFT truncation. This can be seen by comparing the $\Mstar$ limits derived from the EFT approach using truncation (Figs.~\ref{fig:Plot_D5_Limits} and~\ref{fig:Plot_D8_Limits}) to those of the simplified model, recalling $m_V=\Mstar\sqrt{g_f g_{\chi}}$.

\begin{figure}[!htbp]
\centering
\includegraphics[width=1.0\linewidth]{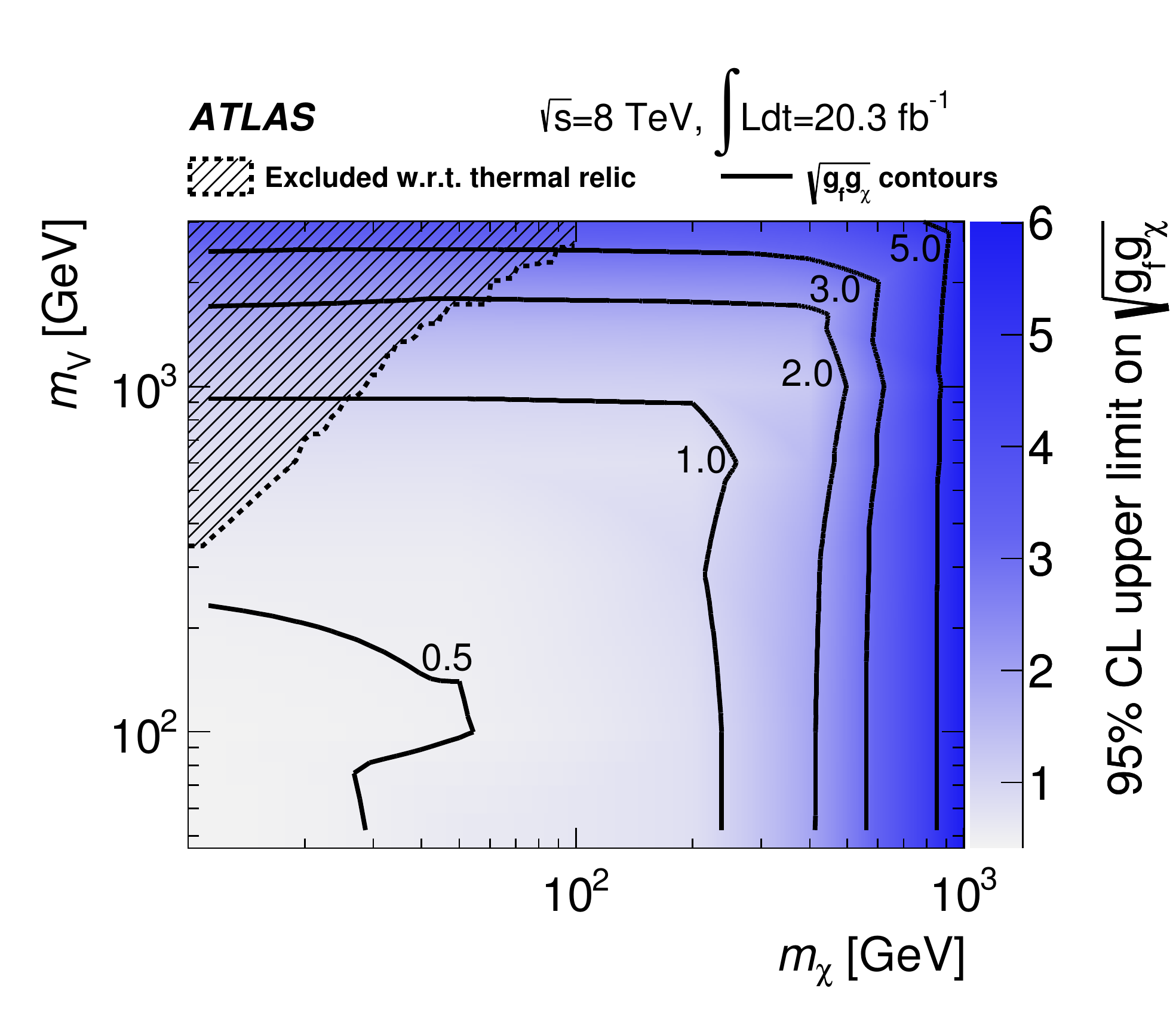}\label{VH_3}
\caption{ Upper limits at  95\% C.L. on the WIMP simplified--model  coupling parameter, $\sqrt{g_fg_\chi}$, with vector coupling and  mediator width $\Gamma=m_V/3$, as a function of the WIMP ($m_\chi$) and the mediator particle ($m_V$) masses.
 Solid lines indicate contours in the coupling parameter.
 The lower limit on the coupling resulting from the relic abundance of DM is also shown.
 }
\label{fig:Plot_VA_Limits}
\end{figure}

\begin{figure}[htbp]\centering
   \includegraphics[width=1.0\linewidth]{{{fig_15}}}
   \caption{ Observed lower limits at
95\% C.L. on the EFT suppression scale $\Mstar$ as a function of the mediator
mass $m_V$, for a $Z'$--like mediator with vector interactions. For a dark--matter
mass $m_\chi$ of $50$ or $400$ GeV, results are shown for different values of the
mediator total decay width $\Gamma$ and compared to the EFT observed limit results
for a D5 (vector) interaction.
 $\Mstar$ vs $m_V$ contours for an overall coupling
$\sqrt{g_fg_\chi}=0.1,0.2,0.5,1,2,5,4\pi$ are also shown.
The corresponding limits from the D5 operator are shown as a  dashed line.
}
   \label{fig:lightM_Lambda_DM50_400_A}
\end{figure}

\begin{figure}[htbp]\centering
   \includegraphics[width=1.0\linewidth]{{{fig_16}}}
   \caption{ Observed limits at 95\% C.L. on the EFT suppression scale $\Mstar$ as a function of the mediator
mass $m_V$, for a $Z'$--like mediator with axial-vector interactions. For a dark--matter
mass $m_\chi$ of $50$ or $400$ GeV, results are shown for different values of the
mediator total decay width $\Gamma$ and compared to the EFT observed limit results
for a D8 (axial-vector) interaction.
 $\Mstar$ vs $m_V$ contours for an overall coupling
$\sqrt{g_fg_\chi}=0.1,0.2,0.5,1,2,5,4\pi$ are also shown.
The corresponding limits from the D8 operator are shown as a dashed line.
}
   \label{fig:lightM_Axial_Lambda_DM50_400_A}
\end{figure}

In the case of the model of $\gamma\gamma\chi\bar{\chi}$ interactions with an $s$-channel SM gauge boson, inspired by the line near 130 GeV in the Fermi-LAT $\gamma$-ray spectrum, limits are placed on the effective mass scale $\Mstar$  in the ($k_2$,$k_1$) parameter plane, as shown in  Fig.~\ref{fig:Plot_schannel_Limit}.
The exclusion line is drawn by considering the value of $\Mstar$ needed to generate the $\chi\bar{\chi}\rightarrow\gamma\gamma$ annihilation rate consistent with the observed Fermi-LAT $\gamma$-ray line near 130 GeV.
This model is able to provide an effective constraint on the portion of the parameter space of the theory compatible with the Fermi-LAT peak.

\begin{figure}\centering
   \includegraphics[width=1.0\linewidth]{{{fig_17}}}
   \caption{ Limits at 95\% C.L. on the effective mass scale $\Mstar$  in the ($k_2$,$k_1$) parameter plane for the $s$-channel EFT model inspired by Fermi-LAT $\gamma$-ray line,   for $m_{\chi} = 130\GeV$.     The upper part of the plane is excluded.
 }
   \label{fig:Plot_schannel_Limit}
\end{figure}

In the ADD model of LED, limits on $M_{\rm D}$ for various values of $n$ are provided in Fig.~\ref{fig:ADD_Limit}. Results incorporating truncation are also shown, for which the graviton production cross section is suppressed by a factor $M_{\rm D}^{4}/\hat{s}^2$, where $\sqrt{\hat{s}}$ is the parton--parton center-of-mass energy.
The analysis is able to exclude $M_{\rm D}$ up to 2.17 \TeV, depending on the number of extra dimensions. The effect of
truncation is larger for higher $n$ as the graviton mass distribution is pushed to higher values.

\begin{figure}[!htbp]
\centering
   \includegraphics[width=1.0\linewidth]{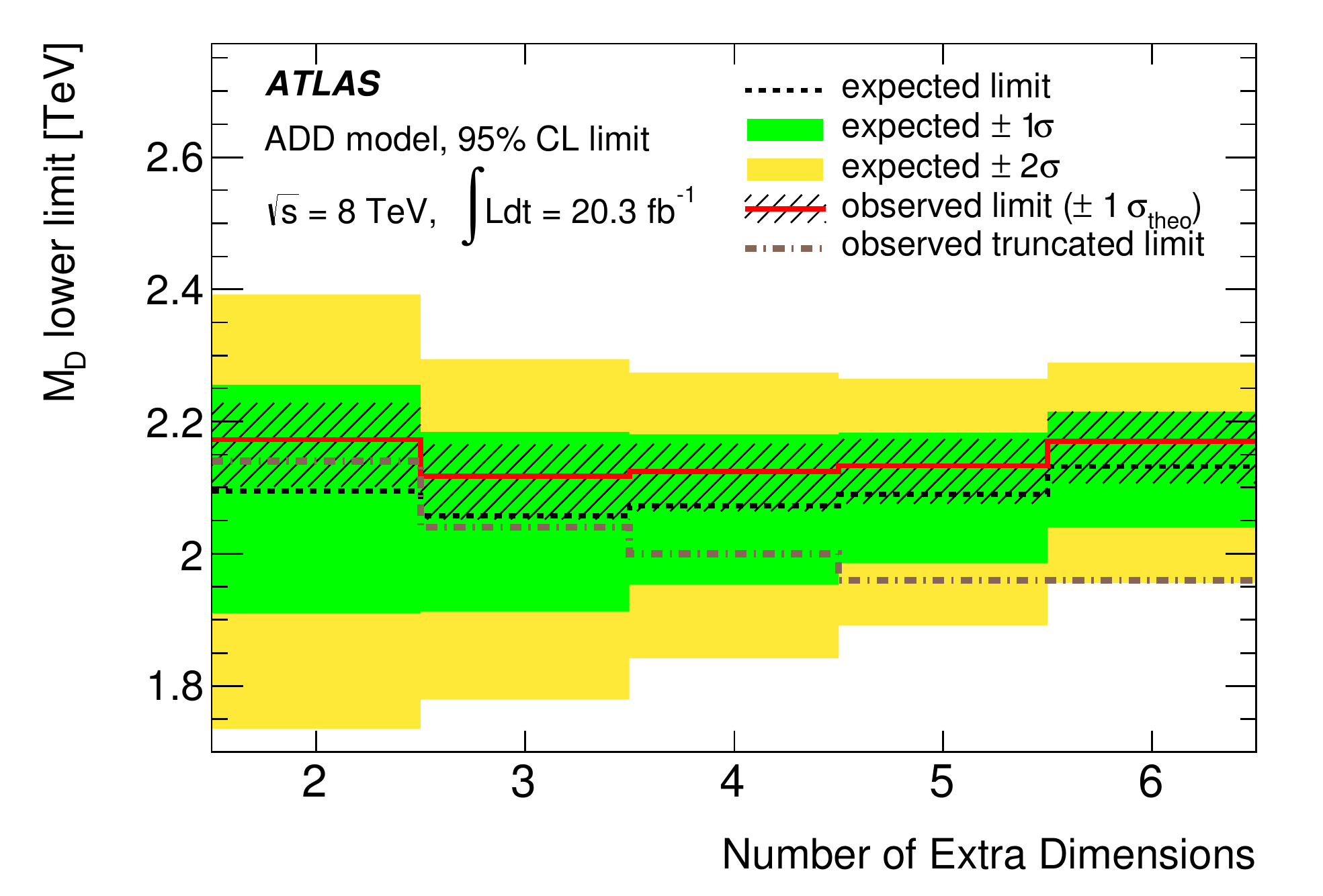}
   \caption{ Lower limits at 95\% C.L. on the mass scale $M_{\rm D}$ in the ADD models of large extra dimensions, for several values of  the number of extra dimensions.
   The expected and observed limits are shown, along with the limit obtained after applying truncation.
   }
   \label{fig:ADD_Limit}
\end{figure}

In the case of squark pair production, limits on $\sigma(pp\rightarrow\tilde{q}\tilde{q}^\ast\gamma+ X)$ as a function of $m_{\tilde{q}}$ and $m_{\tilde{q}}-m_{\tilde{\chi}_{1}^{0}}$  are presented in Fig.~\ref{fig:Plot_Squark_Limit}. The limit is presented  down to $m_{\tilde{q}}-m_{\tilde{\chi}_{1}^{0}}=m_c$, below which the decay of the $\tilde{c}\rightarrow c\tilde\chi_1^0$ is off shell and not considered here.
For very compressed spectra, the analysis is able to exclude squark masses up to 250 \GeV.
Some models of first- and second-generation squark pair production are also explored in Ref.~\cite{Aad:2014wea}; the result presented here is complementary in that it probes very compressed spectra.
Due to the reduced hadronic activity, the acceptance of the $\gamma+\missET$ selection indeed increases as the mass difference between the squarks and
the neutralino decreases, leading to an increased sensitivity to squark mass with decreasing mass difference.

\begin{figure}[!htbp]
\centering
   \includegraphics[width=1.0\linewidth]{{{fig_19}}}
   \caption{Upper limits at 95\% C.L. on the cross section for the compressed squark model, as a function of  the squark mass, $m_{\tilde{q}}$, and of the difference between the squark mass and the mass of the neutralino, $m_{\tilde{q}}-m_{\tilde{\chi}_{1}^{0}}$, in the compressed region of $m_{\tilde{q}} - m_{\tilde{\chi}_{1}^{0}} < 50$ GeV.
The observed (solid line) and expected (dashed line) upper limits from this
analysis are shown; the upper limit on the cross section (in fb) is indicated for each model point.
   }
   \label{fig:Plot_Squark_Limit}
\end{figure}

\FloatBarrier
\section{VIII. Summary}

Results are reported from a search for new phenomena in events with a high-$\pt$ photon and large missing transverse momentum in $pp$ collisions at $\sqrt{s}=8$~TeV at the LHC, using ATLAS data corresponding to an integrated luminosity of 20.3 fb$^{-1}$. The observed data are in agreement with the SM background prediction.  The expected (observed) upper limits on the fiducial cross section $\sigma\times A$ are 6.1 (5.3) fb at 95\% C.L. and 5.1 (4.4) fb at 90\% C.L. In addition, limits are placed on parameters of theories of large extra dimensions, WIMP dark--matter, and supersymmetric quarks.


\section{Acknowledgements}

We thank CERN for the very successful operation of the LHC, as well as the
support staff from our institutions without whom ATLAS could not be
operated efficiently.

We acknowledge the support of ANPCyT, Argentina; YerPhI, Armenia; ARC,
Australia; BMWFW and FWF, Austria; ANAS, Azerbaijan; SSTC, Belarus; CNPq and FAPESP,
Brazil; NSERC, NRC and CFI, Canada; CERN; CONICYT, Chile; CAS, MOST and NSFC,
China; COLCIENCIAS, Colombia; MSMT CR, MPO CR and VSC CR, Czech Republic;
DNRF, DNSRC and Lundbeck Foundation, Denmark; EPLANET, ERC and NSRF, European Union;
IN2P3-CNRS, CEA-DSM/IRFU, France; GNSF, Georgia; BMBF, DFG, HGF, MPG and AvH
Foundation, Germany; GSRT and NSRF, Greece; ISF, MINERVA, GIF, I-CORE and Benoziyo Center,
Israel; INFN, Italy; MEXT and JSPS, Japan; CNRST, Morocco; FOM and NWO,
Netherlands; BRF and RCN, Norway; MNiSW and NCN, Poland; GRICES and FCT, Portugal; MNE/IFA, Romania; MES of Russia and ROSATOM, Russian Federation; JINR; MSTD,
Serbia; MSSR, Slovakia; ARRS and MIZ\v{S}, Slovenia; DST/NRF, South Africa;
MINECO, Spain; SRC and Wallenberg Foundation, Sweden; SER, SNSF and Cantons of
Bern and Geneva, Switzerland; NSC, Taiwan; TAEK, Turkey; STFC, the Royal
Society and Leverhulme Trust, United Kingdom; DOE and NSF, United States of
America.

The crucial computing support from all WLCG partners is acknowledged
gratefully, in particular from CERN and the ATLAS Tier-1 facilities at
TRIUMF (Canada), NDGF (Denmark, Norway, Sweden), CC-IN2P3 (France),
KIT/GridKA (Germany), INFN-CNAF (Italy), NL-T1 (Netherlands), PIC (Spain),
ASGC (Taiwan), RAL (UK) and BNL (USA) and in the Tier-2 facilities
worldwide.

\bibliographystyle{apsrev4-1}

\begin{thebibliography}{87}
\bibitem [{coo()}]{coord}%
  \BibitemOpen
  \href@noop {} {}\bibinfo {note} {ATLAS uses a right-handed coordinate system
  with its origin at the nominal interaction point (IP) in the center of the
  detector and the $z$-axis along the beam pipe. The $x$-axis points from the
  IP to the center of the LHC ring, and the $y$-axis points upward. Cylindrical
  coordinates ($r,\phi$) are used in the transverse plane, $\phi$ being the
  azimuthal angle around the beam pipe. The pseudorapidity is defined in terms
  of the polar $\theta$ angle as ${\eta =
  -\ln\left[\tan(\theta/2)\right]}$.}\BibitemShut {Stop}%
\end{thebibliography}

%

\newpage
\onecolumngrid
\begin{flushleft}
{\Large The ATLAS Collaboration}

\bigskip

G.~Aad$^{\rm 85}$,
B.~Abbott$^{\rm 113}$,
J.~Abdallah$^{\rm 152}$,
S.~Abdel~Khalek$^{\rm 117}$,
O.~Abdinov$^{\rm 11}$,
R.~Aben$^{\rm 107}$,
B.~Abi$^{\rm 114}$,
M.~Abolins$^{\rm 90}$,
O.S.~AbouZeid$^{\rm 159}$,
H.~Abramowicz$^{\rm 154}$
% ...
\bigskip
\\
$^{1}$ Department of Physics, University of Adelaide, Adelaide, Australia\\
% ...
$^{11}$ Institute of Physics, Azerbaijan Academy of Sciences, Baku, Azerbaijan\\
% ...
$^{85}$ CPPM, Aix-Marseille Universit{\'e} and CNRS/IN2P3, Marseille, France\\
% ...
$^{90}$ Department of Physics and Astronomy, Michigan State University, East Lansing MI, United States of America\\
% ...
$^{107}$ Nikhef National Institute for Subatomic Physics and University of Amsterdam, Amsterdam, Netherlands\\
% ...
$^{113}$ Homer L. Dodge Department of Physics and Astronomy, University of Oklahoma, Norman OK, United States of America\\
$^{114}$ Department of Physics, Oklahoma State University, Stillwater OK, United States of America\\
% ...
$^{117}$ LAL, Universit{\'e} Paris-Sud and CNRS/IN2P3, Orsay, France\\
% ...
$^{152}$ Institute of Physics, Academia Sinica, Taipei, Taiwan\\
% ...
$^{154}$ Raymond and Beverly Sackler School of Physics and Astronomy, Tel Aviv University, Tel Aviv, Israel\\
% ...
$^{159}$ Department of Physics, University of Toronto, Toronto ON, Canada\\
% ...
\end{flushleft}


\end{document}